\documentclass[twocolumn,aps,prb,tightenlines,amsmath,amssymb,superscriptaddress]{revtex4}
\usepackage{graphicx}
\usepackage{dcolumn}
\begin{document}
\title{Low-temperature photocarrier dynamics in monolayer MoS$_2$}
\author{T.\ Korn}
\email{tobias.korn@physik.uni-regensburg.de}
\affiliation{Institut
f\"ur Experimentelle und Angewandte Physik, Universit\"at
Regensburg, D-93040 Regensburg, Germany}
\author{S.\ Heydrich}
\affiliation{Institut f\"ur Experimentelle und Angewandte Physik,
Universit\"at Regensburg, D-93040 Regensburg, Germany}
\author{M.\ Hirmer}
\affiliation{Institut f\"ur Experimentelle und Angewandte Physik,
Universit\"at Regensburg, D-93040 Regensburg, Germany}
\author{J.\ Schmutzler}
\affiliation{Institut f\"ur Experimentelle und Angewandte Physik,
Universit\"at Regensburg, D-93040 Regensburg, Germany}
\author{C.\ Sch\"uller}
\affiliation{Institut f\"ur Experimentelle und Angewandte Physik,
Universit\"at Regensburg, D-93040 Regensburg, Germany}
\date{\today}
\begin{abstract}
The band structure of MoS$_2$ strongly depends on the number of layers, and a transition from  indirect to direct-gap semiconductor has been observed recently for a single layer of MoS$_2$. Single-layer MoS$_2$ therefore becomes an efficient emitter of photoluminescence even at room temperature.
Here, we report on scanning Raman and on temperature-dependent, as well as time-resolved photoluminescence measurements on single-layer MoS$_2$ flakes prepared by exfoliation. We observe the emergence of two distinct photoluminescence peaks at low temperatures. The photocarrier recombination at low temperatures occurs on the few-picosecond timescale, but with increasing temperatures, a biexponential photoluminescence decay with a longer-lived component is observed.
\end{abstract}
\maketitle
With the discovery of graphene and the exfoliation technique for preparing single-layer samples from bulk materials, layered crystal structures, in which the binding energy between adjacent planes is much lower than the binding energy within a plane, have attracted a lot of interest in recent years~\cite{Novoselov26072005}. While graphene has many fascinating properties~\cite{Geim_Review}, its lack of a band gap hinders the development of transistors, and the emission of photoluminescence in the visible range has only been observed under pulsed excitation~\cite{Heinz_GraphenePL_PRL10,Wachtrup_PRB10}. The dichalcogenide MoS$_2$, which is used commercially, e.g., as a   lubricant, has been investigated as an alternative to graphene nanoribbons for room-temperature transistor operation~\cite{Kis_NatNano10}, and was recently shown to undergo a transition from indirect to direct-gap semiconductor as its thickness is reduced to a single layer~\cite{Heinz_PRL10,Splen_Nano10}. Similar to graphene, where Raman scattering has been used to determine the layer thickness~\cite{Ferrari_PRL06} or the effects of nanolithography on the carrier concentration~\cite{Steffi_APL10}, Raman spectroscopy of MoS$_2$  is a highly useful tool to identify single layers~\cite{Heinz_ACSNano10}.

Here, we report on Raman scattering and on temperature-dependent, time-resolved photoluminescence measurements on MoS$_2$ flakes. The MoS$_2$ flakes were prepared with the transparent tape liftoff method well-established for graphene, from natural MoS$_2$. A silicon wafer with 300~nm SiO$_2$ layer and lithographically defined metal markers was used as a substrate. After initial characterization with an optical microscope, the samples were analyzed by Raman spectroscopy at room temperature. For this,we utilized a microscope setup, in which a 532~nm cw laser was coupled into a 100x microscope objective, which also collected the scattered light in backscattering geometry. The scattered light was recorded using a triple grating spectrometer equipped with a liquid-nitrogen-cooled charge-coupled device (CCD) sensor. The sample was mounted on a piezo-stepper table and scanned under the microscope. The spatial resolution of this setup is about 500~nm.
For low-temperature photoluminescence (PL) measurements, the sample was mounted in a He-flow cryostat. A  microscope setup with a 40x objective, into which  a 532~nm cw laser was coupled, was used to collect the PL. The spatial resolution of this setup is about 1~$\mu$m. The PL was recorded using a single-grating spectrometer equipped with a CCD sensor. A low pass filter with an onset at a wavelength of 600~nm was used in front of the spectrometer slit to suppress stray light from the laser. Time-resolved PL (TRPL) measurements were performed using the same microscope setup. The second harmonic (wavelength 402~nm) from a picosecond Ti:Sapphire laser was used to excite the sample, the PL was collected with  a streak camera system. The time resolution of this setup is about 5~ps.

\begin{figure}
\includegraphics*[width=\linewidth]{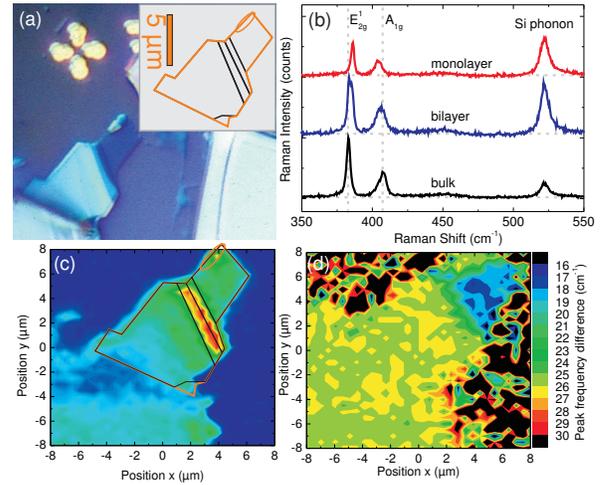}
\caption{(a) Optical micrograph of MoS$_2$ flake on Si/SiO$_2$ substrate. The inset shows the shape of the flake. (b) Normalized Raman spectra measured on different areas of the flake. (c) Intensity map of the E$^1_{2g}$ Raman mode measured on the flake. The flake shape is superimposed on the intensity map. (d) Map of the frequency difference of the E$^1_{2g}$ and A$_{1g}$ Raman modes. The scan area is identical to (c).}
\label{Raman_4Panel}
\end{figure}

First, we discuss the Raman experiments on our sample. Fig.~\ref{Raman_4Panel}(a) shows an optical micrograph of the investigated flake, which shows well-defined steps in the apparent color of the flake, indicating areas which differ in thickness. The inset contains an outline of the flake shape. Typical Raman spectra measured on different regions of the flake are shown in Fig.~\ref{Raman_4Panel}(b). We observe  two characteristic Raman modes of MoS$_2$: the E$^1_{2g}$ mode, which corresponds to an in-plane motion of Mo and S atoms, and the A$_{1g}$ mode, which is due to an out-of-plane vibration of Mo and S. Additionally, the  LO phonon Raman mode of the Si wafer beneath the flake is visible in the spectra, its intensity depends on the thickness of the MoS$_2$. As reported previously~\cite{Heinz_ACSNano10}, the frequency and linewidth of the modes depends on the number of layers of the flake, and  especially the difference of the mode frequencies is a clear indicator for the number of layers. To determine the layer thickness of the different regions of the flake, we performed scanning Raman measurements with a stepsize of 500~nm. The spectra collected for each position on the flake were analyzed using an automated fitting routine which determines the central position and the integrated intensity of the two MoS$_2$ Raman modes and the Si LO phonon mode. The data extracted in this manner was used to generate the false color plots shown in Fig.~\ref{Raman_4Panel}(c) and (d): in Fig.~\ref{Raman_4Panel}(c), we plot the integrated intensity of the E$^1_{2g}$ mode as a function of position. Superimposed on this intensity map is the outline of the flake. We clearly see that the E$^1_{2g}$ intensity closely maps the outline of the flake, a maximum of the intensity is observed in a small region of the flake that separates the near-transparent area from the thicker area of the flake, as seen in the optical micrograph. Since the E$^1_{2g}$ intensity was reported to have a maximum at a  thickness of 4 layers, we may assign that thickness to this region. To further analyze the regions of the flake, we determined the difference of the E$^1_{2g}$ and A$_{1g}$ modes from the Raman scans, as depicted in Fig.~\ref{Raman_4Panel}(d). Here, we see that only the small region on the top right of the flake shows a frequency difference of about 18~cm$^{-1}$, corresponding to a single layer, while the other areas have larger frequency differences between 21~cm$^{-1}$ and 26~cm$^{-1}$, corresponding to thicker layers.

\begin{figure}
\includegraphics*[width=\linewidth]{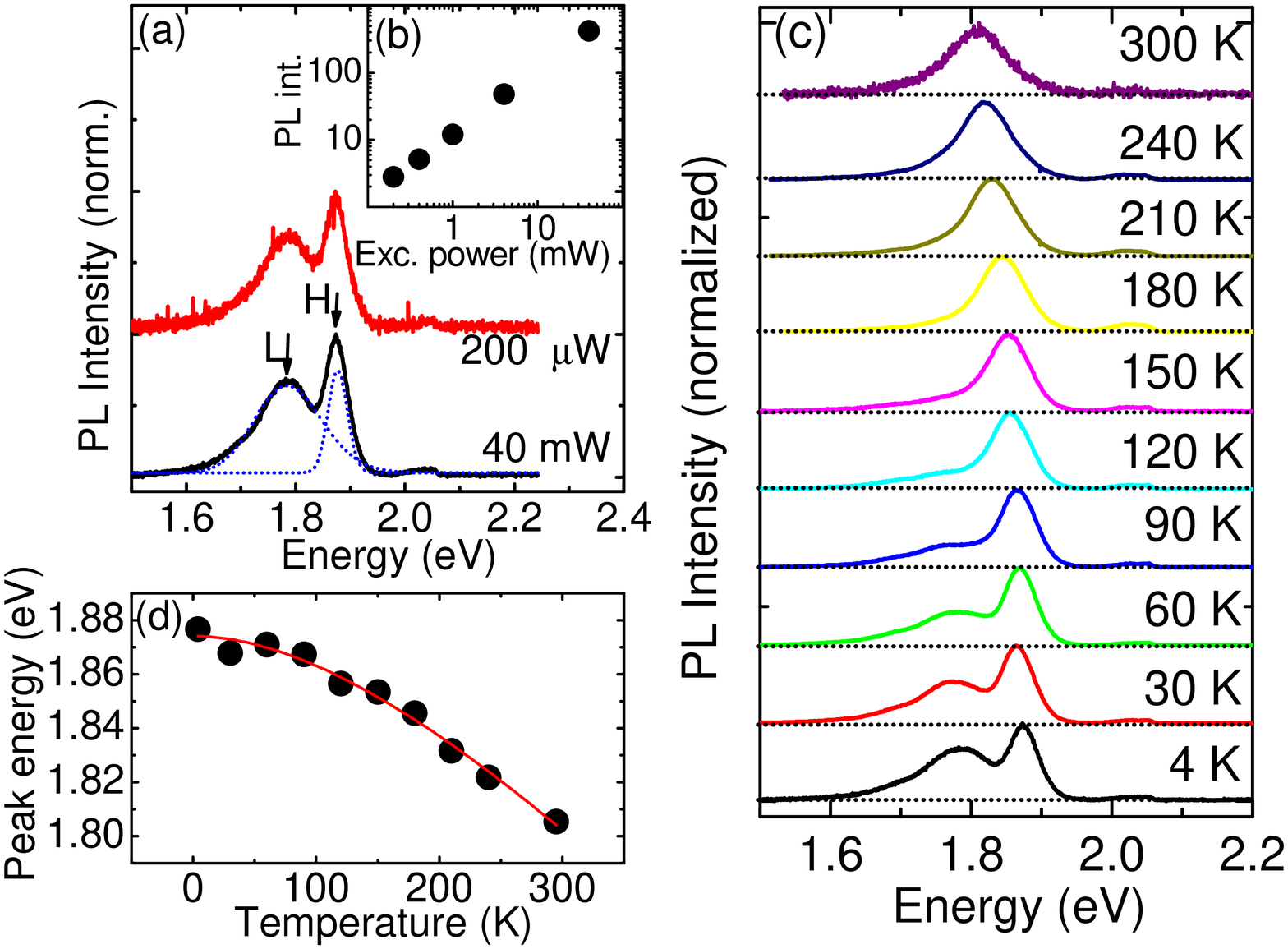}
\caption{(a) Normalized photoluminescence spectra measured at 4.5~K on the monolayer part of the flake for different excitation powers. The two prominent peaks are marked as L and H. In the bottom spectrum, the two Gaussian fit functions to the spectrum are shown. (b) Integrated photoluminescence intensity as a function of excitation power. (c) Normalized PL spectra measured  on the monolayer part of the flake for different temperatures. (d) H peak energy  as a function of temperature (solid dots). The red line indicates a fit using the Varshni equation.}
\label{PL_Power_4K}
\end{figure}

Next, we discuss the PL measurements on the same flake. For these measurements, the excitation laser was focussed onto the portion of the flake identified to be a single layer. Scans of the flake (not shown) confirmed that appreciable PL is only observed from this region. Fig.~\ref{PL_Power_4K}(a) shows typical PL spectra of the single layer region measured at 4.5~K. The spectrum consists of two peaks, which are well-approximated by Gaussian fit functions, a spectrally broad peak at lower energy (marked as L in the figure), and a spectrally narrow peak at higher energy (marked as H). The energy difference $\Delta E$ between these peaks is about 90~meV. The relative intensity of these two peaks does not vary with spatial position on the  single layer region (not shown). Remarkably, the shape of the PL spectrum also remains unchanged as the excitation power is varied by more than two orders of magnitude, as the two spectra in Fig.~\ref{PL_Power_4K}(a) demonstrate. Additionally, the total PL intensity (determined from the integrated area of the two peaks, normalized by the PL exposure time) is proportional to the excitation power in the whole intensity range investigated here (Fig.~\ref{PL_Power_4K}(b)). This indicates that there is neither a saturation of the absorption within the single layer at high excitation powers, nor an appreciable threshold for the emergence of PL due to the presence of defects allowing for nonradiative recombination at low excitation powers. The fact that the relative intensity of the two peaks does not change with excitation power indicates that the states corresponding to the two peaks form independently and that there is no appreciable population transfer from the high-energy state to the other during the photocarrier lifetime.

We now turn to temperature-dependent PL measurements. In this measurement series, two excitation powers were used for each temperature (40~mW and 1~mW). Again, for a fixed temperature, we observe no change of the shape of the PL with excitation power (not shown).  With rising temperature, the two observed peaks broaden and shift to lower energy, as Fig.~\ref{PL_Power_4K}(c) shows. The energy difference $\Delta E$ between peaks L and H remains  constant in the temperature range in which we clearly observe both peaks. We note that the integrated intensity of peak L decreases with temperature, and above 120~K, it is only observable as a low-energy shoulder of peak H. To quantify the spectral shift of peak H with energy, depicted in Fig.~\ref{PL_Power_4K}(d),  we use the  Varshni~\cite{Varshni_Physica67} equation,
$E_g(T)=E_g(0)-\frac{\alpha T^2}{T+\beta},$ which describes the band gap reduction with temperature for many semiconductors. We observe that the energy shift of peak H matches this equation, with $E_g(0)=1.874$~eV, $\alpha = 5.9\cdot 10^{-4}$~eV/K and $\beta =430$~K. We may tentatively identify peaks H and L as the free exciton peak (H) and a bound exciton peak (L), whose energy is reduced by additional binding to defects, either due to impurities at the sample surface or at the MoS$_2$-SiO$_2$ interface. The large width of peak L is an indication that there are binding sites with different energies available for excitons. The quenching of the bound exciton peak with temperature is, most likely, not due to thermal activation, as the binding energy exceeds the thermal energy even at room temperature, but due to an increased probability for nonradiative recombination with temperature.
\begin{figure}
\includegraphics*[width=\linewidth]{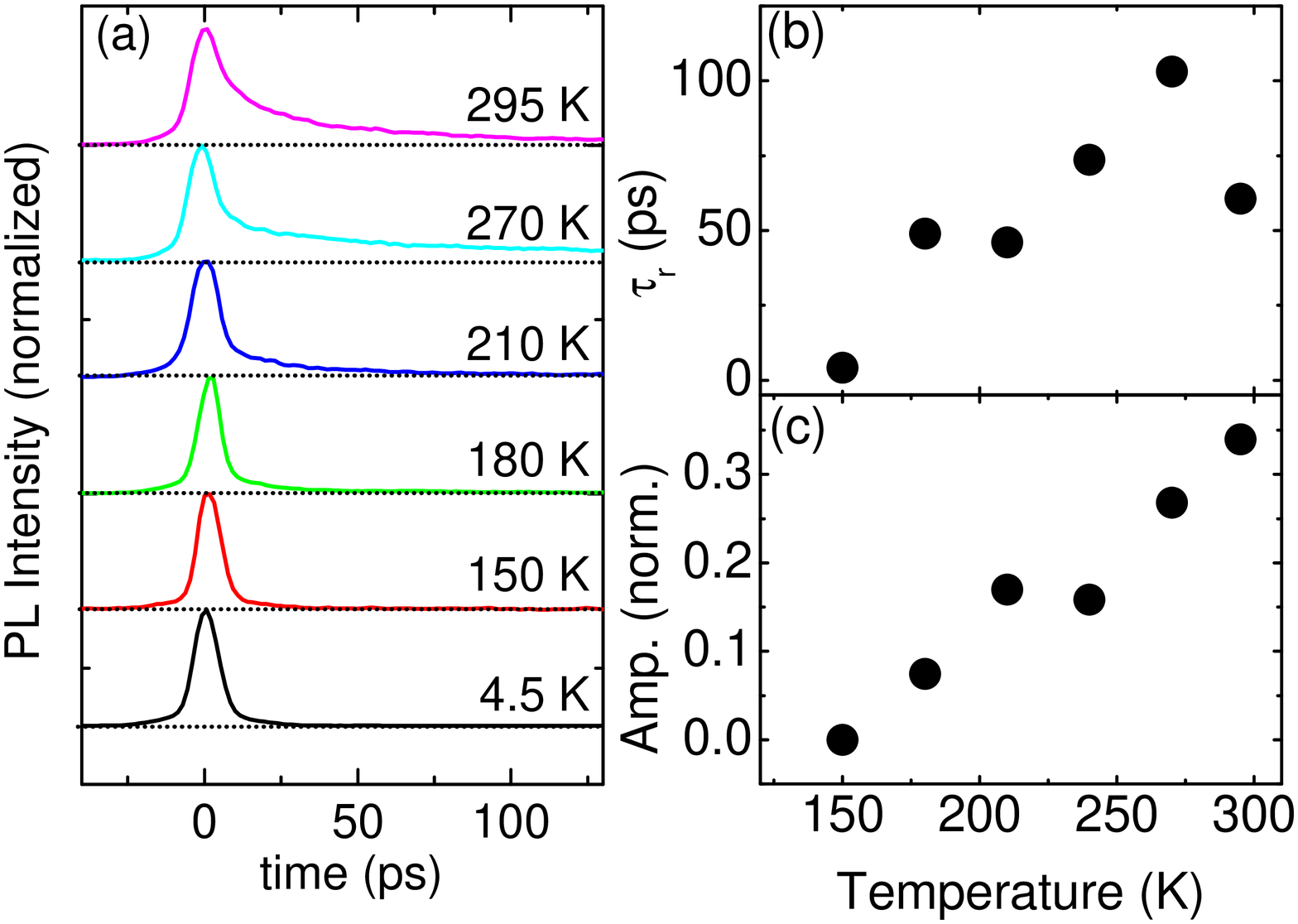}
\caption{(a)Normalized TRPL traces measured  on the monolayer part of the flake for different temperatures. (b) Slow component of photocarrier recombination, $\tau_r$, as a function of temperature. (c) Amplitude of the slow component of the TRPL traces as a function of temperature.}
\label{TRPL_Temp}
\end{figure}
Finally, we discuss the time-resolved PL measurements. Fig.~\ref{TRPL_Temp}(a) shows a series of TRPL traces which were generated by spectrally averaging the TRPL data in a 20~nm wide region around the peak of the PL. In the temperature range from 4.5~K to 150~K,  the PL decays in about 5~ps.  This fast PL decay at low temperatures indicates that there is no full energy relaxation of the optically generated electron-hole pairs into the lowest-energy state (the bound exciton L) during the photocarrier lifetime, as already inferred from the power dependence of the PL. At temperatures above 150~K, we observe that the PL develops a longer-lived component.  In order to extract its time dependence, we fit a biexponential decay function to the data. In Fig.~\ref{TRPL_Temp}(b), we plot the temperature dependence of the  decay time of the long-lived component, $\tau_r$. We observe an increase of this decay time from about 50~ps at 180~K, where the slower decay becomes discernible, to more than 100~ps for 270~K. At room temperature, the decay time decreases again to about 70~ps.  The amplitude of the long-lived PL component increases monotonously with temperature, as  Fig.~\ref{TRPL_Temp}(c) shows. We may attribute the appearance of this long-lived PL to exciton-phonon scattering, which scatters the exciton out of the light cone, i.e. the region in the exciton dispersion where a photon may be emitted while energy and momentum conservation laws are fulfilled~\cite{Shields00}. After such a scattering, the excitons have to reduce their momenta again via subsequent scattering events before they may recombine radiatively.

In conclusion, we have used scanning Raman spectroscopy to identify single-layer MoS$_2$ flakes. We have studied the photoluminescence of single-layer MoS$_2$ with high temporal resolution as a function of temperature. We observe the appearance of a low-energy PL peak at low sample temperatures, which we attribute to a bound exciton state. Time-resolved PL shows that the PL decays on the ps timescale at low temperatures, but develops a long-lived component at higher temperatures, most likely due to exciton-phonon scattering. These observations make MoS$_2$ an interesting material for possible optoelectronic applications, e.g., as a building block for fast photoconductive switches.

The authors gratefully acknowledge financial support by the DFG via SFB689, SPP 1285 and GrK 1570, as well as technical support by J. Eroms and S. Bange. The  MoS$_2$ was supplied by L. Marasz.
\bibliography{MoS2}
\end{document}